\documentclass[preprint2]{aastex}

\newcommand{\lya}{Ly$\alpha$\,\,}
\newcommand{\qa}{PKS\,0528--250}
\newcommand{\qb}{Q\,2206--1958}
\newcommand{\qc}{2233.9+1318}

\shorttitle{Are high-redshift damped Lyman$\alpha$ galaxies
Lyman-break galaxies?}  \shortauthors{M\o ller et al.}

\begin{document}

\title{Are high-redshift damped Lyman$\alpha$ galaxies Lyman-break
galaxies? {}\footnote{\rm Based on observations made with
the NASA/ESA Hubble Space Telescope.}
{}\footnote{\rm Based on observations collected at the European
Southern Observatory, Paranal, Chile (ESO Programme 63.O-0618)}}

\author{P. M\o ller}
\affil{European Southern Observatory, Karl-Schwarzschild-Stra\ss e 2,
D--85748 Garching bei M\"unchen, Germany}
\email{pmoller@eso.org}
\author{S. J. Warren}
\affil{Blackett Laboratory, Imperial College of Science Technology and
Medicine, Prince Consort Rd, London SW7 2BW, UK}
\email{s.j.warren@ic.ac.uk}
\author{S. M. Fall}
\affil{Space Telescope Science Institute, 3700 San Martin Drive,
Baltimore, MD 21218}
\email{fall@stsci.edu}
\author{J. U. Fynbo}
\affil{European Southern Observatory, Karl-Schwarzschild-Stra\ss e 2,
D--85748 Garching bei M\"unchen, Germany}
\email{jfynbo@eso.org}
\and
\author{P. Jakobsen}
\affil{Astrophysics Division, European Space Research and Technology
Centre, 2200 AG Noordwijk, Netherlands}
\email{pjakobsen@astro.estec.esa.nl}

\begin{abstract} 
We use deep HST STIS and NICMOS images of three spectroscopically
confirmed galaxy counterparts of high-redshift damped Ly$\alpha$
(DLA) absorbers (one of which is a new discovery) to test the hypothesis
that high-redshift DLA galaxies are Lyman-break galaxies. If this
hypothesis is correct the emission properties of DLA galaxies must lie
within the range of emission properties measured for Lyman-break
galaxies of similar absolute magnitude. This will be true regardless
of selection biases in the sample of detected DLA galaxies. We test
this prediction using several emission properties: half-light radius,
radial profile (Sersic $n$ parameter), optical-to-near-infrared
colour, morphology, Ly$\alpha$ emission equivalent width, and
Ly$\alpha$ emission velocity structure. In all cases the measured
values for the DLA galaxies lie within the range measured for the
population of Lyman-break galaxies. None of the measurements is in
conflict with the prediction. We conclude that the measured emission
properties of the three DLA galaxies studied here are consistent with
the conjecture that high-redshift DLA galaxies are Lyman-break
galaxies. We show that this result does not conflict with the
observation that the few high-redshift DLA galaxies discovered are
mostly fainter than spectroscopically confirmed $L^*$ Lyman-break
galaxies.  
\end{abstract}


\section{Introduction} 
At any redshift the majority of the neutral hydrogen in the Universe
is contained in the absorbers of highest column density, the damped
Ly$\alpha$ (DLA) absorbers. Observations of the column density
distribution functions $f(N,z)$ for both HI and metals can be used to
compute the cosmic mean densities $\Omega_{\rm HI}$ and $\Omega_m$ as
functions of redshift $z$. The metallicities of the DLA systems have a
large scatter at all redshifts (Pettini et al. 1999; Prochaska \&
Wolfe 1999); the mean metallicity is 5-10\% solar at $z \approx 2-3$
and, within the current statistical uncertainties, may or may not
increase with decreasing redshift, in accordance with models of cosmic
chemical evolution (Kulkarni \& Fall 2002).  Selection biases caused
by dust could dilute the apparent rate of chemical enrichment in the
DLA absorbers (Pei \& Fall 1995; Boiss{\'e} et al. 1998). The
dust-to-metals ratio also has a large scatter among individual DLA
absorbers, but the mean value appears to remain roughly constant with
redshift at about the value in the Milky Way (Pei, Fall, \& Bechtold
1991; Pettini et al. 1997). The presence of metals implies that DLA
absorbers are, or have been, in a state of active star formation and
it is therefore interesting to ask whether those stellar populations
are themselves visible, and if they are related to other known classes
of high redshift objects. In this paper we shall narrowly focus on a
single question: ``Are high-redshift damped Ly$\alpha$ galaxies
Lyman-break galaxies?''.

The Lyman-break galaxies (LBGs, Steidel et al. 1996) are starburst
galaxies at high redshift, recognisable by their relatively flat
restframe ultraviolet continuum, and a sharp discontinuity at the
Lyman limit. Under the definition of LBGs we include all objects
identified by their characteristic continuum shape as such, on the
basis of broadband photometry alone i.e. without requiring
spectroscopic confirmation. The typical brightness of an LBG is not
a well defined quantity as it depends strongly on the flux limit of
the survey. In particular one may consider two distinct types of LBG
survey, where surveys concerned solely with the determination of
photometric redshifts have significantly fainter flux limits than
those which classify as LBGs only galaxies with spectroscopically
confirmed redshifts. It is of interest to note that to an AB magnitude
limit $H=26.5$, i.e. flux limited in the restframe optical continuum,
nearly all galaxies in the Hubble Deep Field N in the redshift range
$2<z<3.5$ are classified as LBGs (Papovich, Dickinson, \& Ferguson
2001). With the caveat that this sample is small, it may be concluded
that for observations in the optical and near-infrared, to very faint
magnitudes, any galaxies with steep continuum, either due to reddening
or to the presence of a population of old stars, are a minority
population. So at these wavelengths, and down to the AB magnitude
limit $H=26.5$, the term `Lyman-break galaxy' is essentially
synonymous with `high-redshift galaxy'.

The global rate of star formation at these redshifts measured from
emission from the LBGs is consistent with the rate of star formation
inferred from the change with cosmic time of $\Omega_{\rm HI}$ and
$\Omega_m$ in the DLA absorbers (Pei, Fall, \& Hauser 1999), modulo
uncertain corrections in both calculations for the effects of
dust. The simplest interpretation of these results is that the DLA
absorbers are the reservoirs of gas from which the stars in LBGs are
forming. If this is the case, for every high-redshift DLA absorber
detected in the spectrum of a quasar there should be stellar emission
visible from a Lyman-break galaxy at the absorption redshift,
coincident with or near to the line of sight to the quasar
\footnote{The covering factor of DLA absorbers is greater than the
fraction of sky covered by detectable optical emission from LBGs. So the
absorbing gas would extend beyond the optically visible stellar
emission.}. Yet searches for this stellar emission have had very
little success (for reviews and discussions of past surveys see
M{\o}ller \& Warren 1993; Kulkarni et al. 2000). The large number of
unsuccessful attempts to identify the galaxy counterparts of DLA
absorbers, hereafter ``DLA galaxies'', indicates that most DLA
galaxies at $z>2$ have very small impact parameters or are very faint
(e.g. Steidel, Pettini, \& Hamilton 1995a). Indeed searches that have
produced lists of candidates pending spectroscopic confirmation
(e.g. Ellison et al.  2001a), as well as the very limited set of
confirmed identifications available at present (Fynbo, M\o ller, \&
Warren 1999), both suggest that DLA galaxies mostly are fainter than
$L^*$ LBGs, a result supported by upper limits on H$\alpha$ emission
(Bunker et al.  1999; Kulkarni et al. 2001). This is seen by some to
contradict the view that DLA galaxies are Lyman-break galaxies. In
particular the theoretical prediction that the angular momentum
distribution of dark matter halos should cause LBGs and DLA galaxies
to form two ``quite distinct populations'' (Mo, Mao, \& White 1999)
has received much attention, and it has been suggested that DLA
galaxies make up a separate low surface brightness galaxy population
(Jimenez, Bowen, \& Matteucci 1999).

At redshifts $z \approx 0$ the DLA galaxies are much easier to identify,
but it is far from clear how they are related to high-redshift DLA
galaxies. Nevertheless it has been found that also the low redshift
DLA galaxies typically are sub $L^*$ (Steidel et al. 1995b;
Lanzetta et al. 1997; Miller, Knezek, \& Bregman 1999; Cohen 2001;
Turnshek et al. 2001; Bouch{\'e} et al. 2001).

In this paper we present results from deep images of three
spectroscopically confirmed DLA galaxies observed with the HST
instruments STIS and NICMOS, and make a comparison of their emission
properties with the emission properties of LBGs. The images presented
here are the first combined (both STIS and NICMOS) results from a
large campaign of HST observations searching for 18 DLA galaxies
towards 16 quasars. The targets, the strategy, and the goals are
presented in detail in our first paper (Warren et al. 2001) which also
presented the results of the NICMOS imaging campaign. Follow-up
spectroscopy is being carried out with the ISAAC and FORS instruments
on the ESO VLT.  In \S2 we present a discussion of how to use such a
search for DLA galaxies to test the hypothesis that DLA galaxies are
LBGs. In \S3 we describe briefly the HST observations, and the
measurements of the images. In \S4 we make a comparison of the
measured properties, including sizes, colours, and morphologies, with
the properties of LBGs {\em of similar absolute magnitude}. In \S5 we
address the question posed at the top, and present our
conclusions. Cosmological parameters of ${\mathrm\Omega_{\circ}=0.3}$,
${\mathrm\Lambda_{\circ}=0.7}$, $h={\mathrm H_{\circ}/100}$ are
assumed, and the term high redshift implies in this paper $2<z<3.5$.

\section{Interpreting searches for DLA galaxies}
Because of the glare from the quasar the detectability of a DLA galaxy
is dependent both on the brightness of the galaxy and the impact
parameter (in addition to the brightness of the quasar and the
stability of the point spread function (psf)). Any sample of DLA
galaxies, including the small sample of three discussed here, is
biased towards brighter objects and larger impact parameters. {\em Yet
if DLA galaxies are LBGs, the measured emission properties
of any DLA galaxy must fall within the range of emission properties
for LBGs of the same absolute magnitude.} This statement is true
regardless of how the DLA galaxies were selected. This is the basic
prediction that we test in this paper. We compare the measured
half-light radii, and colours of DLA galaxies, as well as the
morphologies, and Ly$\alpha$ emission properties, against the same
properties measured for LBGs of similar absolute magnitude.

If DLA galaxies are LBGs they will nevertheless have a different
luminosity distribution compared to a flux limited sample of LBGs,
because DLA absorbers are detected in proportion to their gas cross
section. It follows that the luminosity distribution of DLA galaxies
is given by the LBG luminosity function weighted by the luminosity
dependence of the gas cross section. A simple calculation illustrates
the importance of this for the interpretation of searches for DLA
galaxies. For a luminosity function of Schechter form, and a power-law
relation between gas radius and luminosity $R\propto L^{t}$, the mean
luminosity of galaxies selected by gas cross section is
$\bar{L}=L^*\Gamma(2+\alpha+2t)/\Gamma(1+\alpha+2t)$ (Fynbo, M\o ller,
\& Warren 1999; see also Wolfe et al. 1986; Impey \& Bothun 1989).
For LBGs at $z\sim3$, Adelberger \& Steidel (2000) measure
$\alpha=-1.57$. Depending on the value of $t$, the value of the mean
luminosity can be much less than $L^*$. For example for $t=0.4$,
$\bar{L}=0.23 L^*$. In fact there is some theoretical support for a
value of $t\sim0.4$ (Haehnelt, Steinmetz, \& Rauch 2000), based on the
kinematics of DLA absorbers as measured from the absorption profiles
of low ionisation species. So the notion that DLA galaxies are
typically fainter than $L^*$ is perfectly compatible with the
hypothesis that DLA galaxies are Lyman-break galaxies. We note that a
consequence of this is that the value of any property which depends on
galaxy luminosity will differ between galaxy samples selected either
by absorption (DLA galaxies) or by emission (LBG galaxies). For
example, since spatial clustering is weaker for less massive (hence less
luminous) galaxies (the effects of natural biasing in a hierarchical
clustering scenario), we expect DLA galaxies to have a lower clustering
amplitude than the emission-selected LBGs.

In the longer term, when we have completed our HST survey, we will
have a sample of DLA galaxies complete within well defined selection
criteria. Using the measured impact parameters we will be able to
determine the form of the relation between radius and luminosity. In
fact we have already attempted this (Fynbo et al. 1999). One point to
note is that the conclusion that DLA galaxies are much fainter than
$L^*$ is quite robust, and insensitive to the uncertainty of the
faint-end slope $\alpha$ of the Lyman-break galaxy luminosity
function.  For a different value of $\alpha$, the value of $t$ must be
adjusted in order to reproduce the observed small impact parameters of
DLA galaxies. The result is that the value of $\bar{L}$ changes
little, because the exponent $\alpha+2t$ appearing in the expression
for $\bar{L}$ remains approximately constant.

\section{Properties of three DLA galaxies}
In this section we detail the measurement of the emission properties
of the three DLA galaxies, which will be compared against the
LBGs in the next section.

\subsection{Observations and targets} 
The larger imaging campaign is described by Warren et al. (2001). We
are obtaining high-resolution optical and near-infrared images of
quasars, searching for faint galaxies close to the quasar line of
sight.  These are candidate DLA galaxies, targets for spectroscopic
follow up. In Warren et al. (2001) we list the 16 target quasars and
their coordinates, we tabulate details of the 18 DLA absorbers,
present details of the NICMOS observations, and the list of NICMOS
candidates. Each quasar was observed for three orbits with the NIC2
camera and the F160W filter. The NICMOS images reach to an AB
magnitude typically of $H_{160}=25$.  A total of 41 candidate
counterparts were detected in boxes of side $7.5 \arcsec$ centered on
each quasar. Spectroscopy of this candidate list is $25\%$ complete.
So far one new spectroscopically confirmed DLA galaxy has been
discovered. The other two DLA galaxies discussed here are previous
discoveries.

In M\o ller et al. (in preparation) we will present the candidates
detected using STIS. With STIS we are imaging in the 50CCD
configuration i.e. without a filter in order to reach as deep as
possible. In this configuration the effective central wavelength is
5851.5\AA\ and the FWHM is 4410.3\AA\, so 50CCD corresponds to a very
wide V magnitude. Each quasar was observed for two orbits, at different
orientations, resulting in a detection limit on the combined image
of typically $V_{50}=27$ (AB magnitude). Because the STIS images reach
much fainter magnitudes than the NICMOS images, the number of candidates
in the STIS images is typically a factor 2-3 higher than in the
NICMOS images.

\begin{figure}
\plotone{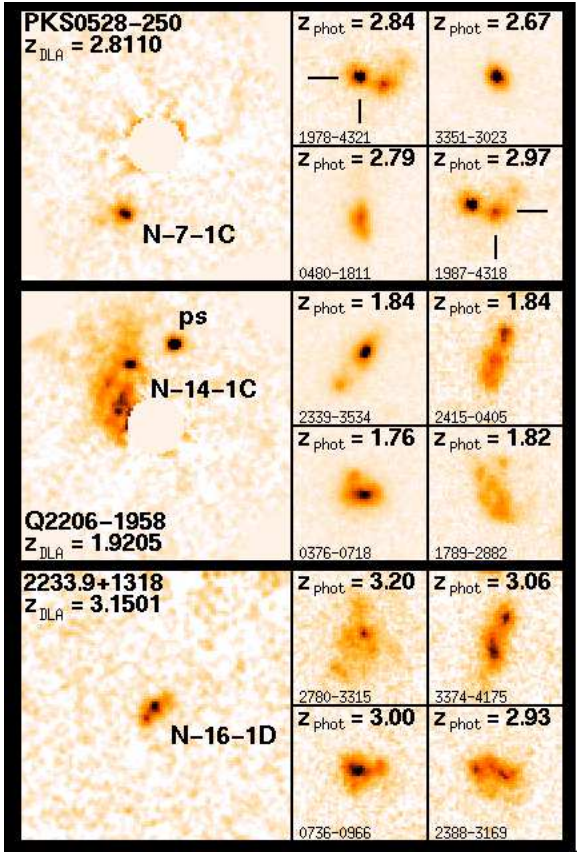}
\figcaption[f1.eps]{Images of the three DLA galaxies, and of
LBGs of similar redshift and luminosity. The three
left-hand frames, of side 4\arcsec, show the images after subtraction
of the quasar psf. The DLA galaxies are marked using the numbering
system from Warren et al. (2001). The top two frames are centred on
the quasar, and strong residuals from the psf subtraction have been
set to the sky level near the quasar centre. The object next to
N-14-1C marked {\it ps} is a red point source, presumably unrelated
to the DLA galaxy. The lowest left-hand
frame is centred on the DLA galaxy, and the quasar is located outside
the frame. To the right of each DLA galaxy image are four frames,
{2\arcsec} on a side, showing LBGs in HDF S, selected, as
described in the text, to have similar redshifts and magnitudes as the
corresponding DLA galaxy.}
\end{figure}

Fig.~1 reproduces the STIS images of the fields of the three quasars
\qa, \qb, and \qc, where the frames from both orbits have been
registered and summed. The images show the field after subtraction of
the quasar image. Strong residuals near the center of the quasar image
have been masked. The DLA galaxies are indicated in each frame,
labeled using the numbering scheme in the NICMOS paper. The measured
impact parameters are $1.14\arcsec$, $0.99\arcsec$, $2.51\arcsec$,
respectively. In each case the confirmed galaxies are the candidates
in the NICMOS frames closest to the line of sight to the quasar.  The
three DLA galaxies have all been confirmed by spectroscopic detection
of \lya in emission, but were originally found with three different
search techniques. N-7-1C was identified on a deep \lya narrow band
image obtained from the ground, N-14-1C was found on the image from
our current STIS campaign (Fig. 1), and is also detected in the NICMOS
image, and N-16-1D was found on ground based images with the
Lyman-break technique. Careful psf-subtraction was in all cases
instrumental in the discovery. We now briefly summarise previous
results obtained on each of the three DLA galaxies.

\subsubsection{N-7-1C} The galaxy N-7-1C was discovered by M\o ller \&
Warren (1993), who called it S1. The DLA absorber lies at a similar
redshift to the quasar.  Using spectroscopic data Warren \& M{\o}ller
(1996) and M\o ller, Warren, \& Fynbo (1998) argued that the physical
separation is sufficiently large that the ionising flux from the quasar
is not important. This result was supported by Ge et al. (1997) who
concluded that the distance between the DLA absorber and the quasar must
be larger than 1 Mpc, on the basis of ionization modelling (see also
Ledoux et al. 1998; Ellison et al. 2001b).  M\o ller \& Warren (1998)
obtained WFPC2 images of N-7-1C and showed that the luminosity profile
is similar to the profile of LBGs at similar redshift.  Ge et al.
(1997) reported a metallicity of the DLA absorber of 10\% solar and a
dust-to-gas ratio of 8\% of the Milky Way value. Lu, Sargent, \&
Barlow (1997) were not able to fit the absorption spectrum of this DLA
absorber into the rotating-disk model of Prochaska \& Wolfe
(1997, 1998), which requires a `leading-edge asymmetry'.

\begin{figure}
\plotone{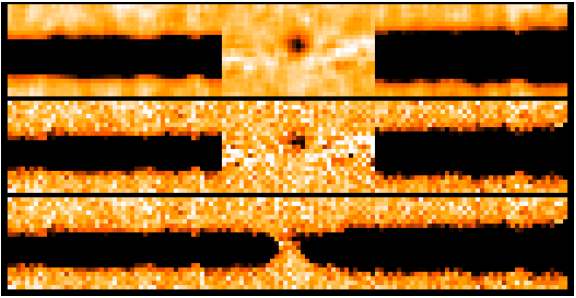}
\caption{Frames showing the 2D spectrum of Q2206-1958
and the DLA galaxy N-14-1C that confirms the galaxy as the counterpart
of the DLA absorption line. Wavelength increases to the right. The
bottom panel shows the summed 8000 second exposure after subtraction
of sky, where the thick dark line is the quasar spectrum, and the gap
is the DLA absorption line. In the middle panel a section of quasar
spectrum centred in wavelength on the DLA absorption line has been
subtracted using the SPSF software (M\o ller, 2000). The DLA galaxy
\lya emission line is visible just above the centreline of the quasar
spectrum and slightly redshifted ($z=1.9229$) relative to the
absorption line ($z=1.9205$). In the top panel the same data have
been smoothed slightly to improve the contrast of the DLA galaxy \lya
emission line over the noise.}
\end{figure}

\subsubsection{N-14-1C}
The galaxy N-14-1C is a new discovery. The spectrum confirming the
identification is provided in Fig.~2. Details are presented in Tables
1 and 2. The metallicity of the corresponding absorber is 1/3 of the
solar value, but despite being one of the most metal rich DLA
absorbers known there is no evidence for dust (Prochaska \& Wolfe
1997).  In this case Prochaska \& Wolfe (1997) found evidence for the
leading edge asymmetry in the low ion absorption lines, which they
interpret as evidence for a rotating disk.

\begin{deluxetable}{llccccrcccc}
\tabletypesize{\scriptsize}
\tablecaption{Photometric and structural properties of three
high-redshift DLA galaxies} 
\tablewidth{0pt}
\tablehead{
 \multicolumn{1}{c}{Quasar} & \multicolumn{1}{c}{Galaxy} &
 \colhead{$\mathrm{V_{50}}$} & \colhead{$\mathrm{H_{160}}$} &
 \multicolumn{2}{c}{$b$\tablenotemark{a}} & \multicolumn{2}{c}{$r_{1/2}$} &
 \colhead{Sersic $n$} & \colhead{Ellip.}  & \colhead{$\phi$\tablenotemark{b}} \\
 & & & & \arcsec & \multicolumn{1}{c}{kpc} & \arcsec & \multicolumn{1}{c}{kpc} & & & deg.
} 
\startdata
\qa & N-7-1C  & $25.43\pm0.11$ & $25.18\pm0.22$ & 1.14 & $6.3h^{-1}$  & $0.09\pm0.01$ &
 $0.5h^{-1}$  & $0.56$  & 0.26 & 74 \\
\qb & N-14-1C & $24.69\pm0.07$ & $24.69\pm0.19$ & 0.99 & $5.8h^{-1}$  & $0.5\pm0.1$ &$3.0h^{-1}$ & \nodata & \nodata & \nodata \\
\qc & N-16-1D & $25.75\pm0.12$ & $25.34\pm0.17$ & 2.51 & $13.3h^{-1}$ & $0.14\pm0.02$ &
 $0.7h^{-1}$  & $0.22$  & 0.61 & 24 \\
 \enddata

\tablenotetext{a}{Impact parameter measured in the STIS image
($\sigma_b = 0.02$\arcsec )}
\tablenotetext{b}{$\phi$ is the angle between the
galaxy orientation and the line joining the galaxy to the quasar
i.e. a galaxy pointing at the quasar has $\phi = 0$}
\tablecomments{For N-14-1C magnitudes are calculated from the
integrated flux in an aperture of 0.6\arcsec\ diameter, with no aperture
correction. All other magnitudes are total magnitudes, computed as
described in the text. All magnitudes are on the AB system.}
\end{deluxetable}

\begin{deluxetable}{lcccccccc}
\tabletypesize{\scriptsize}
\tablecaption{\lya emission line properties of three high-redshift DLA
galaxies}
\tablewidth{0pt}
\tablehead{
 \multicolumn{1}{c}{Galaxy} & \colhead{$z_{\mathrm{qso}}$} &
 \colhead{$\mathrm{log(N_{HI})}$} &
 \colhead{$z_{\mathrm{DLA}}$} & \colhead{$z_{\mathrm{Ly\alpha}}$} &
 \colhead{$\Delta v$ (\lya--DLA)} & \colhead{\lya FWHM} & \colhead{\lya
 flux} & \colhead{\lya EW$_{\rm rest}$} \\ 
 && cm$^{-2}$ & & & km s$^{-1}$ & km s$^{-1}$ & erg s$^{-1}$ cm$^{-2}$ &
 $\mathrm{\AA}$  
} 
\startdata
N-7-1C & 2.797 & 21.35 & 2.8110 & 2.8136(5) & $+190\pm 40$ & $680\pm 75$ &
          $7.4\pm 0.6 \times 10^{-17}$ & $63\pm 8$ \\
N-14-1C& 3.559 & 20.65 & 1.9205 & 1.9229(6) & $+250\pm 50$ & $600\pm 200$ &
          $2.6 \pm 0.3 \times 10^{-16}$ & $83\pm 11$ \\
N-16-1D& 3.298 & 20.00 & 3.1501 & 3.1530(3) & $+210\pm 30$ & 360 &
          $6.4 \pm 1.2 \times 10^{-17}$ & $82\pm 19$ \\
 \enddata

\tablecomments{The data are taken from: this work, Djorgovski et
al. (1996), Lu, Sargent and Barlow (1997), Warren and M\o ller (1996),
M\o ller and Warren (1998), Warren et al. (2001)}
\end{deluxetable}

\subsubsection{N-16-1D}
The galaxy N-16-1D was first reported by Steidel, Pettini, \& Hamilton
(1995a) who called it N1. Spectroscopic confirmation of the redshift
was obtained by Djorgovski et al. (1996). Lu, Sargent, \& Barlow
(1997) reported a metallicity `typical of DLA galaxies at such
redshifts' ([Fe/H] = -1.4). They also found evidence for leading-edge
asymmetry in the absorption spectrum.

\subsection{Results} {\it Photometry and profile fitting.}  
In Table 1 we summarise the results of STIS and NICMOS photometry, and
profile fitting of the galaxies. The columns list successively the
quasar name, the galaxy name, the STIS and NICMOS AB magnitudes, the
galaxy impact parameter $b$, i.e. the angular offset of the galaxy
from the quasar, and the details of the deconvolved surface-brightness
profiles measured on the STIS frames. The measurement of N-14-1C was
problematic, and is described later.  For N-7-1C and N-16-1D we
followed the same procedure described by Warren et
al. (2001). Briefly, the best fit galaxy profile is determined by
convolving parameterised models with the point spread function, and
searching for the $\chi^2$ minimum.
The $68\%$ confidence intervals on each parameter,
marginalising over the other parameters, are computed on the basis of
the Poisson errors for each pixel by locating the orthogonal tangent
planes to the $\Delta \chi^2=1.0$ ellipsoid. We used the Sersic model,
where the surface brightness as a function of radius $r$ is
$\Sigma=\Sigma_{1/2} \exp\{-B(n)\lbrack(r/r_{1/2})^{1/n}-1\rbrack\}$,
where $r_{1/2}$ is the deconvolved half-light radius. Note that $n=1$
corresponds to an exponential profile, and $n=4$ to a de Vaucouleurs
profile (for details see Warren et al. 2001).
Table 1 lists $r_{1/2}$, the value of $n$, the galaxy
ellipticity, and the parameter $\phi$ which is the angle between the
galaxy major axis and the line pointing to the quasar.  For these two
sources the STIS magnitudes are total magnitudes, calculated by
extrapolating the model fits to infinite radius. Because of the low
$S/N$ of the NICMOS frames total NICMOS magnitudes were measured by
determining the scaling of the STIS models that provided the best fit
to the NICMOS data.

The galaxy N-14-1C has an irregular morphology (Fig.~1) with a bright
knot at the end of a diffuse structure. The measured parameters of the
fit varied greatly with the size of the fitting region. Therefore we
resorted to aperture photometry to measure magnitudes, with a circular
aperture of 0.6\arcsec\, diameter. In order to obtain an estimate of the
half-light radius we employed a model with fewer parameters, a
circularly symmetric exponential model, which was much less sensitive
to box size. Since the model is not a good fit we assigned a large
uncertainty, $0.1\arcsec$, to this measurement.

{\it Spectroscopy.}  As stated above the galaxy N-14-1C is a new
discovery. The confirming spectrum was obtained with the FORS1
instrument on the European Southern Observatory 8.2m telescope UT1 on
the night of 1999 August 12. The integration time was 8000s.  With a
slit width of $1.3\arcsec$ and the 600B grating the resolution was
3.0\AA. We used the SPSF method of M\o ller (2000) to extract the
galaxy spectrum. The software optimises the spectral extraction of a
faint galaxy close to a bright companion (here the quasar). In Fig. 2
we show the region of the spectrum around the wavelength of the DLA
absorber, $z_{DLA}=1.9205$. An emission line at a wavelength of
3553.4\AA, corresponding to \lya at redshift of $z=1.9229\pm 0.0006$,
is clearly visible. The close agreement between the emission and
absorption redshifts, and the small impact parameter of the galaxy
confirm that N-14-1C is the DLA absorber galaxy counterpart.

In Table 2 we summarise relevant properties of the \lya emission line
of the three galaxies. In successive columns are listed the galaxy name,
the quasar redshift, the column density and redshift of the absorption
line, and then the following properties of the emission lines: redshift,
velocity difference relative to the absorber, FWHM, flux, and rest-frame
equivalent width.  

\section{Comparison with other high-redshift galaxies}
In this section we compare the measured emission properties for the
three DLA galaxies against the distribution of emission properties
measured for LBGs, also observed with HST. We begin by considering the
half light radii and the observed optical-to-near-infrared colours. To
obtain a comparison sample of LBGs we started with the catalogue of
Cohen et al. (2000) of redshifts for galaxies in the Hubble Deep Field
(HDF) N. We selected the 26 galaxies with spectroscopic redshifts in
the range $2.0<z<3.5$. This sample was reduced to 24 by removing two
galaxies where the published redshifts are the subject of debate
(Papovich et al. 2001). Marleau \& Simard (1998) have measured the
surface brightness profiles of galaxies in the HDF using Simard's
GIM2D software. We matched 21 of the 24 galaxies with their catalogue,
and this forms our comparison sample of LBGs.

\begin{figure}
\vskip -2.0cm
\plotone{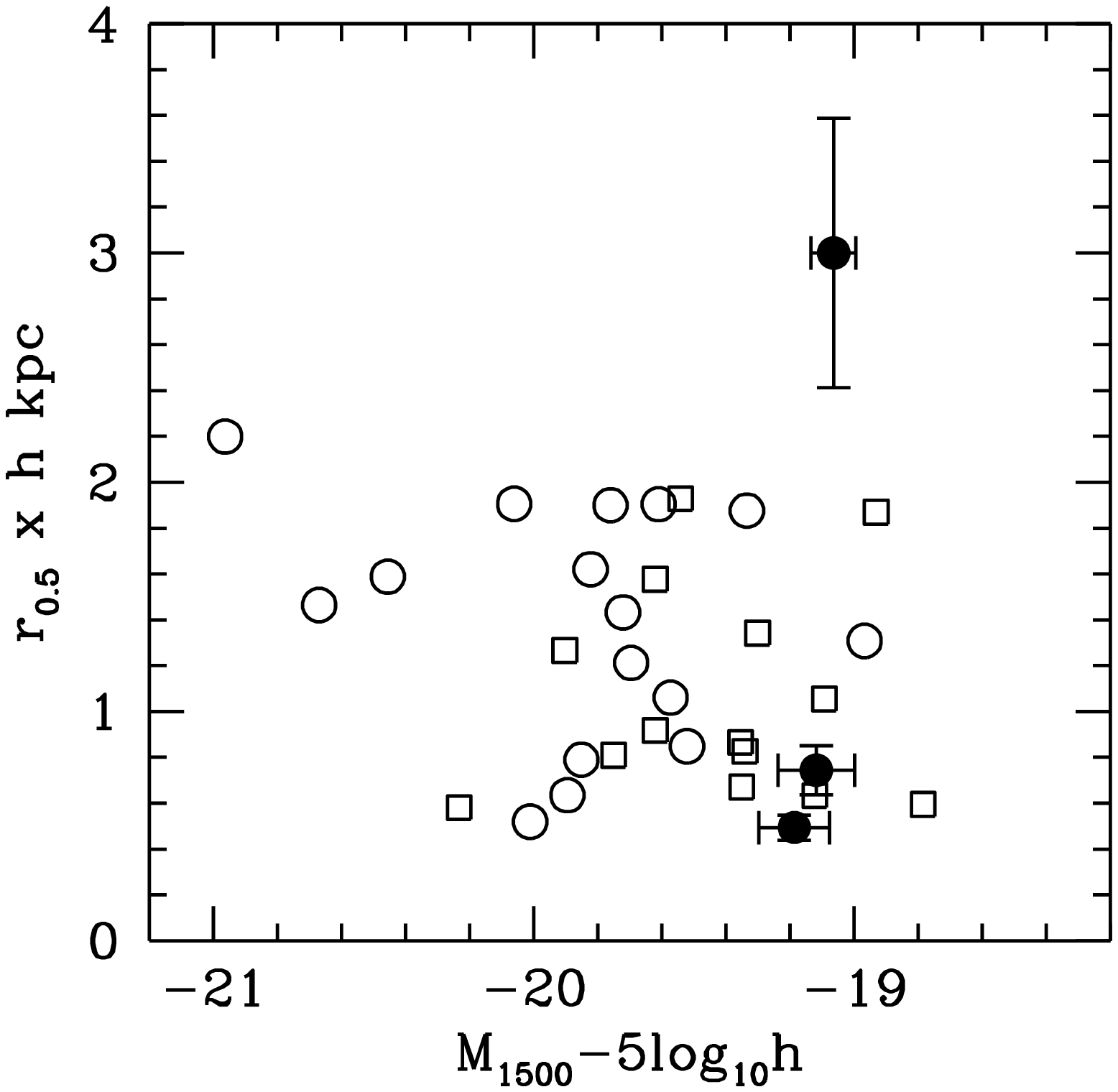}
\vskip -1.6cm
\plotone{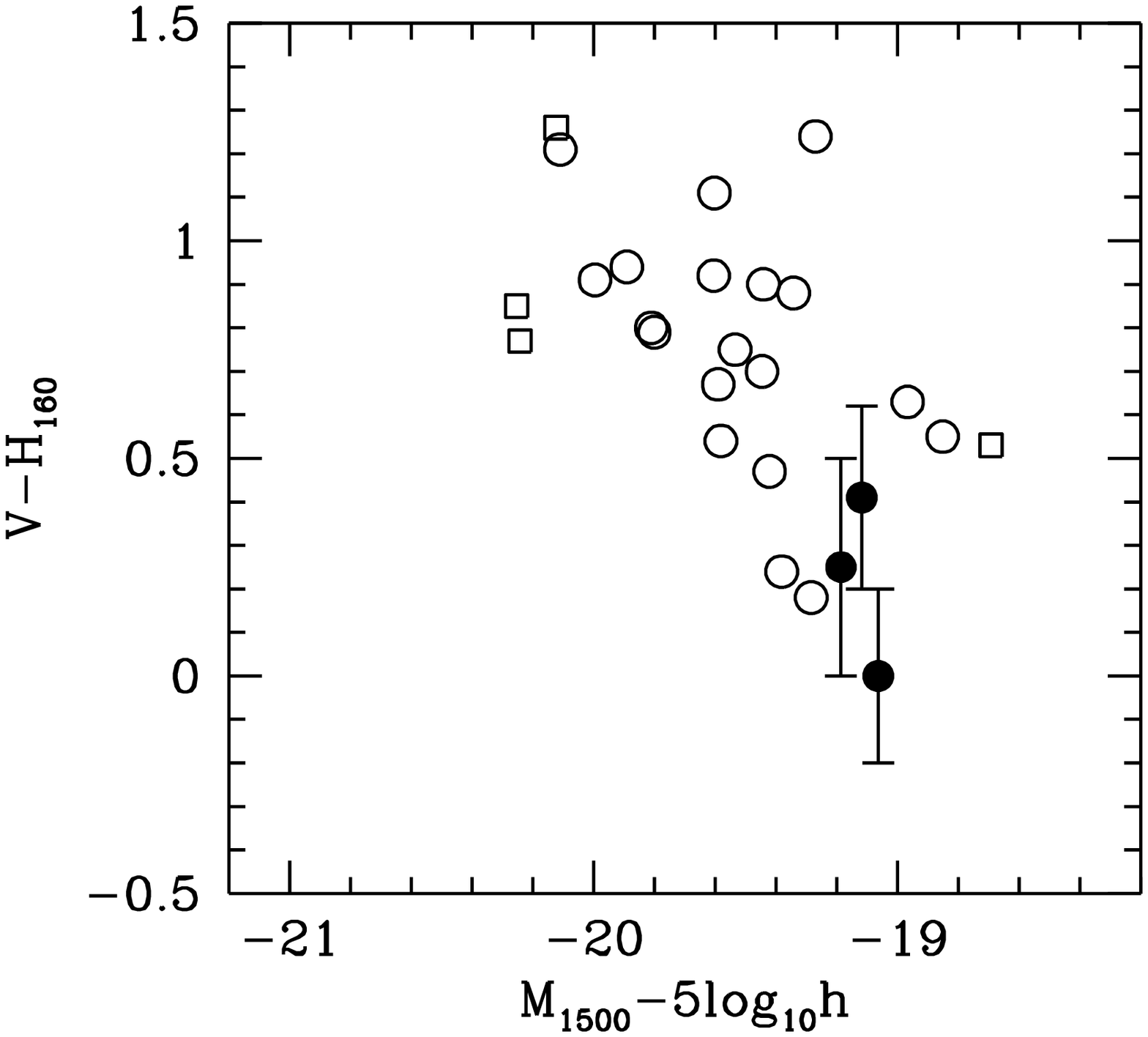}
\vskip -0.6cm
\caption{{\it Upper plot: } Plot of half-light
radius against absolute magnitude for 21 HDF N LBGs
$2.0<z<3.5$, measured by Marleau and Simard (1998), and the three DLA
galaxies. The 16 LBGs consisting of a single component are plotted as
open circles. The 14 components of the remaining five LBGs are plotted
as open squares. The three DLA galaxies are plotted as filled circles.
The DLA galaxies have half-light radii in the same range as the LBGs
of similar absolute magnitude. {\it Lower plot: }
Plot of optical-to-near-infrared $V-H_{160}$ colour against absolute
magnitude for the same 21 LBGs and three DLA galaxies
from the upper plot, and using the same symbols. The colours of the
LBGs are taken from Papovich et al. (2001). Their
catalogue differs from that of Marleau and Simard (1998) in terms of
the number of galaxy components, with 19 single galaxies plus two
galaxies with a total of four components. Therefore we have used the
$V_{606}$ magnitudes of Papovich et al. to calculate absolute
magnitudes. The DLA galaxies have similar colours to the Lyman-break
galaxies of similar absolute magnitude.}
\end{figure}

{\it Half-light radii:} In the list of Marleau \& Simard (1998) 16 of
the 21 galaxies show a single component. The other five galaxies are
decomposed into a total of 14 components. We extracted from their
catalogue the half light radius $r_{1/2}$ and the total apparent AB
magnitude for the WFCAM F606W passband, $V_{606}$, for each of the 30
galaxies or components. (Note that the pivot wavelength for the WFCAM
F606W configuration is very close to the pivot wavelength for the STIS
50CCD passband.) These data are presented in Fig.~3 (upper plot),
plotting half-light radius against absolute magnitude, $M_{1500}$ (the
AB absolute magnitude at a rest wavelength $\lambda=1500$\AA\ \---\ we
computed the small $k$ corrections assuming a continuum shape
$f_{\nu}\propto \nu^{-0.5}$). The LBGs are plotted as open symbols.
Our measurements of the same parameters for the three DLA galaxies are
plotted as filled symbols. There is no strong trend of size with
luminosity for the LBGs. The DLA galaxies have similar luminosities to
the faintest of the LBGs plotted. The range of half-light radii for
the DLA galaxies is similar to the range for the LBGs, although the
half-light radius for N-14-1C is a little larger than the LBG radii
plotted. Note, however, that it is quite likely that Marleau and
Simard would have decomposed a galaxy like N-14-1C into two
components.

Marleau \& Simard (1998) found that the LBGs mostly
have small bulge-to-disk ratios. The small values of the Sersic
parameter $n$ listed in Table 1 for N-7-1C and N-16-1D imply similarly
disk-dominated profiles. At present we cannot quantify how N-14-1C
fits into this pattern.

{\it Optical-to-near-infrared colours:} Papovich et al. (2001) provide
optical-to-near-infrared ($V_{606}-H_{160}$) colours for high-redshift
galaxies in the HDF N. For the same 21 galaxies considered above they
list 23 components. Since there is not a one-to-one correspondence
between their list of components and the list of Marleau \& Simard
(above) we have used the total apparent magnitudes from Papovich et
al. to compute absolute magnitudes (their total magnitudes are
typically 0.3 mag. fainter than the corresponding values in Marleau
\& Simard 1998).

In Fig.~3 (lower plot) we plot $V-H_{160}$ colour against absolute
magnitude for the LBGs and DLA galaxies, using the same symbols
as in the upper plot. As noted by Papovich et al. (2001) there is a
marked trend of colour with luminosity, in the sense that galaxies of
lower luminosity are bluer. They point out that this trend could
reflect a sequence in age, in metallicity, or in dust content, but
they do not draw any conclusions on this point. The DLA galaxies are
very faint and blue, and have colours in the same range as LBGs of
the same absolute magnitude.

{\it Morphology:} In addition to the size/luminosity relation, one may
consider the more detailed morphology of the objects. Are DLA galaxies
for example more disturbed, or do they have more components than
LBGs? To address this question we searched the list of
Fontana et al. (2000) for LBGs in the HDF S with
redshift and V magnitudes as similar as possible to each of the three
DLA galaxies. To be specific we selected, for each DLA galaxy, the
four LBGs with $V_{606}$ and $z_{phot}$ that gave the
smallest value of the sum $(V_{606} - V_{50,DLA})^2 + (z_{phot} -
z_{DLA})^2$. We then extracted a $2\arcsec\times2\arcsec$ image
centred on each object. The four objects are shown in Fig.~1 next to
the corresponding DLA galaxy. The three DLA galaxies themselves span a
wide range in general morphology. N-7-1C has a regular single core
morphology, N-14-1C is dominated by a single bright knot at the end of
an extended structure, and N-16-1D is elongated with two or three
compact components. Qualitatively, one can find examples that resemble
each of the DLA galaxies amongst the Lyman-break galaxy images.

{\it \lya emission/absorption:} The restframe line equivalent widths
(EWs) for the Ly$\alpha$ emission lines from the DLA galaxies N-7-1C and
N-16-1D are given in the literature as 51 and 37 \AA, but with large
errors due to uncertainties in determining their broad band fluxes. We
have here calculated EWs for all three DLA galaxies as follows: The \lya
line falls in all cases inside the 50CCD passband, so we first subtract
the measured \lya flux from the $V_{50}$ flux. A linear interpolation
of the $H_{160}$ and $V_{50}$ (corrected for \lya) fluxes then allow a
computation of the continuum flux at the wavelength of the redshifted
\lya line. The resulting EWs are listed in Table 2, and cover the range
63 to 83\AA. The median EW found here is larger than that for LBGs
(about 50\% of the LBGs have \lya in absorption so their median EW is
close to zero). The distribution of Ly$\alpha$ EWs for confirmed DLA
galaxies will inevitably be skewed towards large emission line EWs as
this makes spectroscopic confirmation easier. Nevertheless the values
reported here lie within the measured range for LBGs (EW$_{\rm abs}$ =
70\AA\ to EW$_{\rm em}$ = 100\AA\ in a field 92\% completed for objects
having $R \leq 25.0$, Steidel et al. 2000), and so are in agreement
with the basic prediction we are testing.

Table 2 also lists the velocity difference between the \lya
emission and absorption lines $\Delta v$ (\lya--DLA) for the three DLA
galaxies. Part of the velocity difference could be due to relative
motion, either in the form of random motion of sub-clumps, of ongoing
merging processes (Haehnelt, Steinmetz, \& Rauch 1998) or of rotation
(Prochaska \& Wolfe 1997). However, it is noticeable that in all cases
the emission line is redshifted relative to the DLA absorber. The lines
are also rather broad $300-600$ km s$^{-1}$ FWHM. The same
characteristics have been noted in the LBGs (e.g. Pettini et al. 2000;
Pettini et al. 2001; see also Schaye 2001) and can be explained in
terms of radiative transfer effects of \lya photons escaping through
outflowing gas. Photons at the \lya wavelength in the rest-frame
of the moving HI are scattered repeatedly and are eventually absorbed by
dust; those in the wings of the \lya line have a smaller scattering
cross-section and escape from the neutral gas (Kunth et al. 1998). In
terms of \lya line emission, the properties of all three DLA galaxies
are therefore in good agreement with those of the LBGs.

\section{Discussion and conclusions}

In this paper we have presented new STIS imaging data and VLT
spectroscopic data on a small set of three DLA galaxies. Two of those
were previously known, one is a new discovery. The small set of three
objects does not allow a conclusive answer to the dispute over whether
DLA absorbers arise in large disks or merging sub-clumps. Our data set
does allow us, however, to address the more fundamental (and
model independent) question of the possible relationship
between DLA and Lyman-break galaxies.

We have compared the emission properties of the three DLA galaxies
with the emission properties of the population of LBGs. If DLA
galaxies are LBGs then the measured values for an emission property
for any DLA galaxy will lie within the range of values measured for
LBGs of similar absolute magnitude \---\ regardless of the selection
biases that are inherent in the sample of DLA galaxies.  We have
tested this prediction, making the comparison for the following
emission properties: half-light radius, radial profile (Sersic $n$
parameter), optical-to-near-infrared colour, morphology, Ly$\alpha$
emission EW, and Ly$\alpha$ emission velocity structure. In all cases
the measured values for the DLA galaxies lie within the range measured
for LBGs. None of the measurements is in conflict with the prediction.
{\em Therefore we conclude that the measured emission properties of the
three damped Ly$\alpha$ galaxies studied here are consistent with the
conjecture that damped Ly$\alpha$ galaxies are Lyman-break galaxies.}

Is the number of detections in the follow-up spectroscopy so far
completed of the 41 NICMOS candidates also consistent with this
conjecture? Since two of the DLA galaxies discussed here were
previously known they are ignored in this statistical analysis. We
have obtained deep VLT FORS spectra, average integration time 6000s,
of 11 of the 22 NICMOS candidates in 6 fields, targeting 8 DLA
absorbers. We have identified and confirmed one new DLA galaxy, and we
would expect to have confirmed $(8\times11\times F_{NIC} \times
F_{Ly\alpha})/22=4\times F_{NIC} \times F_{Ly\alpha}$ new DLA
galaxies. Here the factor $F_{NIC}$ is the fraction of DLA galaxies
detectable in the NICMOS images i.e. above the magnitude and
impact-parameter detection limits of our survey (approximately
$H_{160}<25$, $b>0.6\arcsec$ \---\ full details provided in Warren et
al., 2001) and $F_{Ly\alpha}$ is the fraction of DLA galaxies with
detectable Ly$\alpha$ flux (our $5\sigma$ detection limit ranges from
0.2 to $2.0 \times 10^{-16}$ erg s$^{-1}$ cm$^{-2}$ depending on
redshift, column density and impact-parameter of the DLA). If DLA
galaxies are LBGs then both of the factors $F_{NIC}$ and $F_{Ly\alpha}$
are likely to be substantially less than unity (\S2
and \S4). Therefore the detection of only one new DLA galaxy so far
is consistent with the hypothesis, but the number of detections is
clearly too small to rule out other interpretations.

The combination of information on both the stellar (emission) and
interstellar (absorption) components of the DLA galaxies is crucial
for a complete understanding of these objects. The result presented
here is a step towards combining the information from absorption-line
and emission studies of high-redshift galaxies. To prove that DLA
galaxies are LBGs, the test we have made is a necessary but not
sufficient condition. We are building a larger sample of DLA galaxies
with well-defined selection properties, from spectroscopy of
candidates found in deep HST images. A definitive test will be
possible with this sample. Knowing the selection biases we can
determine the relation between gas radius and luminosity that explains
the number of DLA galaxies detected, their luminosities, and their
impact parameters.

\acknowledgments

We thank Martin Haehnelt for a helpful conversation and Max Pettini,
Jason Prochaska, and Hsiao Wen Chen for helpful comments on an earlier
version of this manuscript. SJW is grateful to ESO for hospitality
during part of the writing of this paper. Part of the reduction of the
STIS data was done at STScI, Baltimore and PM gratefully acknowledges
support from the cheerful STScI support staff and in particular useful
discussions with Charles Proffitt on the subject of STIS psfs. SMF is
grateful to ESO and the Max-Planck Institut f{\"u}r Astrophysik,
Garching for hospitality in the early stages of this work and the
Carnegie Observatories, Pasadena, in the late stages.

\end{document}